\documentclass{ws-ijmpd}

\usepackage{url}

\begin{document}


%
\catchline{}{}{}{}{}
%

\title{Multi-messenger astronomy with Centaurus A}

\author{M.~Kachelrie{\ss}}
\address{Institutt for fysikk, NTNU, Trondheim, Norway}

\author{S.~Ostapchenko}
\address{Institutt for fysikk, NTNU, Trondheim,
  Norway\\ D.~V.~Skobeltsyn Institute of Nuclear Physics, 
Moscow State University, Russia}

\author{R.~Tom\`as\footnote{Speaker}}
\address{II. Institut f\"ur Theoretische Physik,
    Universit\"at Hamburg, Germany}

\maketitle

\begin{abstract}
We calculated for the nearest active galactic nucleus (AGN),
Centaurus~A (Cen A), the flux of high energy cosmic rays and of accompanying
secondary photons and neutrinos expected from hadronic interactions in
the source.  We used as two basic models for the generation of
ultrahigh energy cosmic rays (UHECR) shock acceleration in the radio
jet and acceleration in the regular electromagnetic field close to the
core of the AGN, normalizing the UHECR flux to the
observations of the Auger experiment. Here we compare the previously
obtained photon fluxes with the recent data reported by the Fermi LAT and
H.E.S.S.\ collaborations. In the case of the core model, we find good agreement 
both for the predicted spectral shape and  the overall normalization 
between our earlier results and the H.E.S.S.\ observations for a primary
proton energy $dN/dE\propto E^{-\alpha}$ with $\alpha\sim 2$ or smaller.
A broken-power law with high-energy part $\alpha=-2.7$ leads to photon
fluxes in excess of the Fermi measurements. The energy spectrum of the
photon fluxes obtained by us for the jet scenario is in all cases at variance 
with the H.E.S.S.\ and Fermi observations.
\end{abstract}


\section{Introduction}
Progress in cosmic ray (CR) physics\cite{CR} has been hampered for long 
time by the deflection of charged cosmic rays in magnetic fields, preventing 
the identification of individual sources. 
This problem could be solved by using the neutral
messengers that should be produced as secondaries in hadronic CR interactions
close to the source.  However, the secondary
photons generated  are difficult to
disentangle from photons produced by synchrotron radiation or inverse
Compton scattering of electrons. Moreover, high energy photons are
strongly absorbed both in the source and propagating over
extragalactic distances. By contrast, the extremely large mean free
path of neutrinos together with the relatively poor angular resolution
of neutrino telescopes  and the small expected event
numbers makes the identification of extragalactic sources challenging
using only the neutrino signal. Performing neutrino astronomy beyond
the establishment of a diffuse neutrino background requires therefore
most likely additional input, either timing or angular information
from high energy photon or CR experiments.

The recently announced evidence\cite{pao_corr} for a correlation of
the arrival directions of UHECRs observed by the Pierre Auger
Observatory (PAO) with active galactic nuclei (AGN) may provide a
first test case for successful ``multi-messenger astronomy.'' In
particular, Ref.~\refcite{pao_corr} finds two events within the search
bin of $3.1^\circ$ around the nearest active galaxy, Centaurus~A
(Cen~A).  This FR~I radio galaxy is located close to the supergalactic
plane at a distance of about 4 Mpc (see Ref.~\refcite{Israel:1998ws}
for more details).  At present
this correlation, not confirmed by the HiRes experiment\cite{HiRes}, has
only $3\,\sigma$~C.L. and other source types that follow the
large-scale structure of matter would also result in an excess of
events along the supergalactic plane.

In Ref.~\refcite{Kachelriess:2008qx}, we studied therefore the possibility
to observe Cen~A using high-energy photons and neutrinos.
Taking the correlation signal at face value, we used the PAO results
as normalization of the CR flux and calculated the flux of accompanying
secondary photons and neutrinos expected from hadronic interactions in
the source.  Since both the Fermi LAT collaboration~\cite{Abdo:2009wu} and 
the H.E.S.S.\ collaboration~\cite{Aharonian:2009xn} reported recently  
the discovery  of $\gamma$-ray emission from Cen~A, we have now the opportunity
to check our predictions against these observations.

\section{Assumptions}

We calculated the flux of high energy cosmic rays and of
accompanying secondary photons and neutrinos expected from Cen~A for
two different scenarios: Acceleration close to the core, either in
accretion shocks or regular electromagnetic fields, and acceleration
in the radio jet. In the first case UV photons are the most important
scattering targets.  We modeled the primary photon field around
the AGN core guided by the simplest possible theoretical
model\cite{sunyaev}, namely the thermal emission from a geometrically thin,
optically thick Keplerian accretion disc. The interaction depth for
photo-hadron interactions can reach $\tau_{p\gamma}\sim $~few.
According to the observational data we found that pp interactions of
UHE protons with the gas provide the main source of CR interactions
in the second scenario.  In this case, moreover, diffusion in the
turbulent magnetic fields will increase the interaction depth at low
and intermediate energies.
We considered  also three spectra ${\rm d}N/{\rm d}E\propto
E^{-\alpha}$ of the 
injected protons: Power-laws with $\alpha=1.2$ and $\alpha=2$, and 
a broken power-law with $\alpha = 2.7$ for $E>E_b=10^{18}\,$eV. 

We based our calculations on several simplifying assumptions like the use 
of an one-dimensional geometry and the omission of the acceleration process.
In particular, we only postulated that acceleration to $10^{20}$\,eV  is 
possible in the 
environment of Cen~A, without demonstrating it for a concrete model.
Finally, hadronic  interactions are simulated with an extension of the Monte 
Carlo code described in Ref.~\refcite{II}.

\section{Results versus observational data}

\begin{figure}[h!]
\begin{center}
\epsfig{file=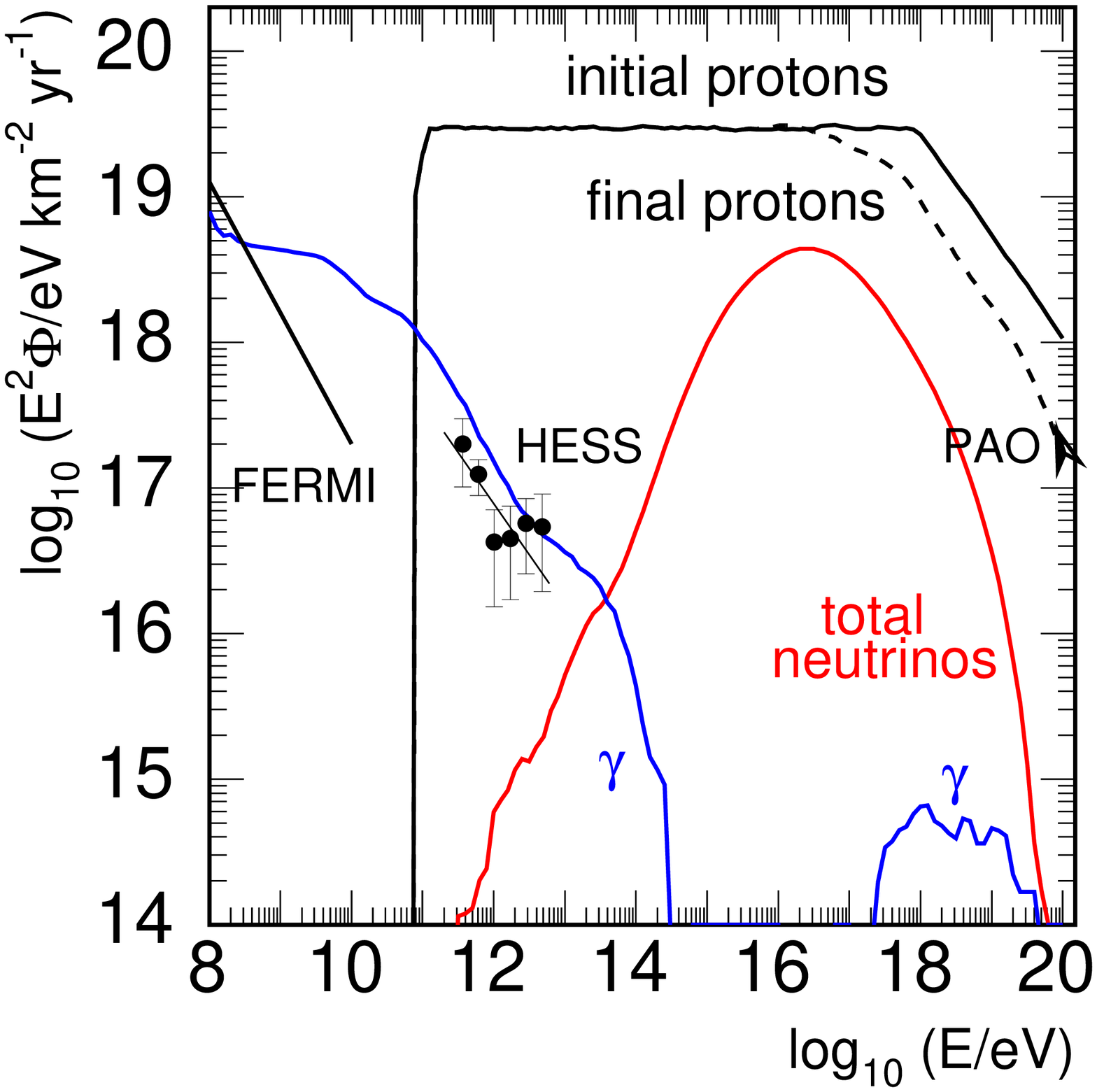,width=0.44\textwidth,angle=0}
\epsfig{file=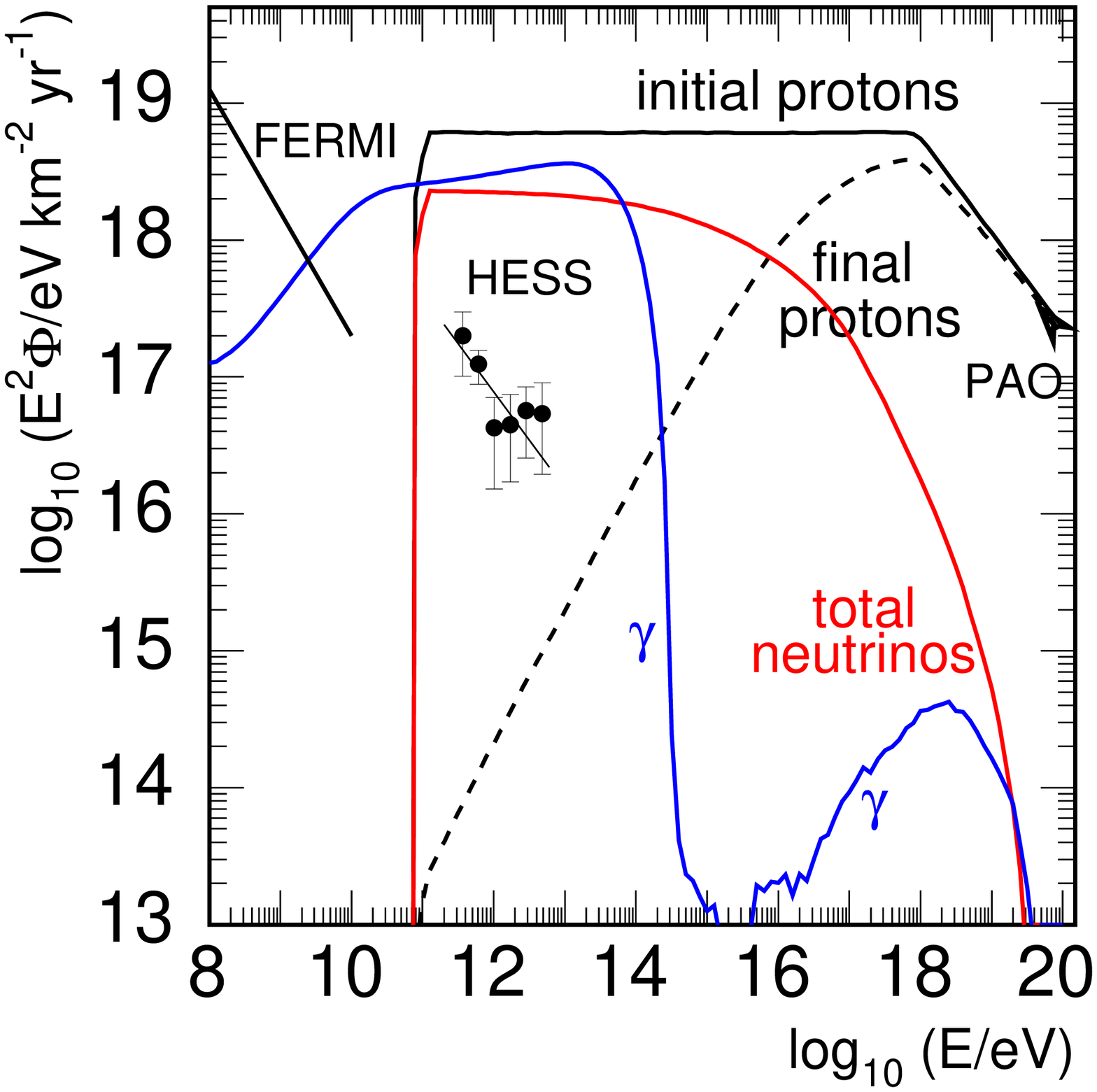,width=0.44\textwidth,angle=0}
\epsfig{file=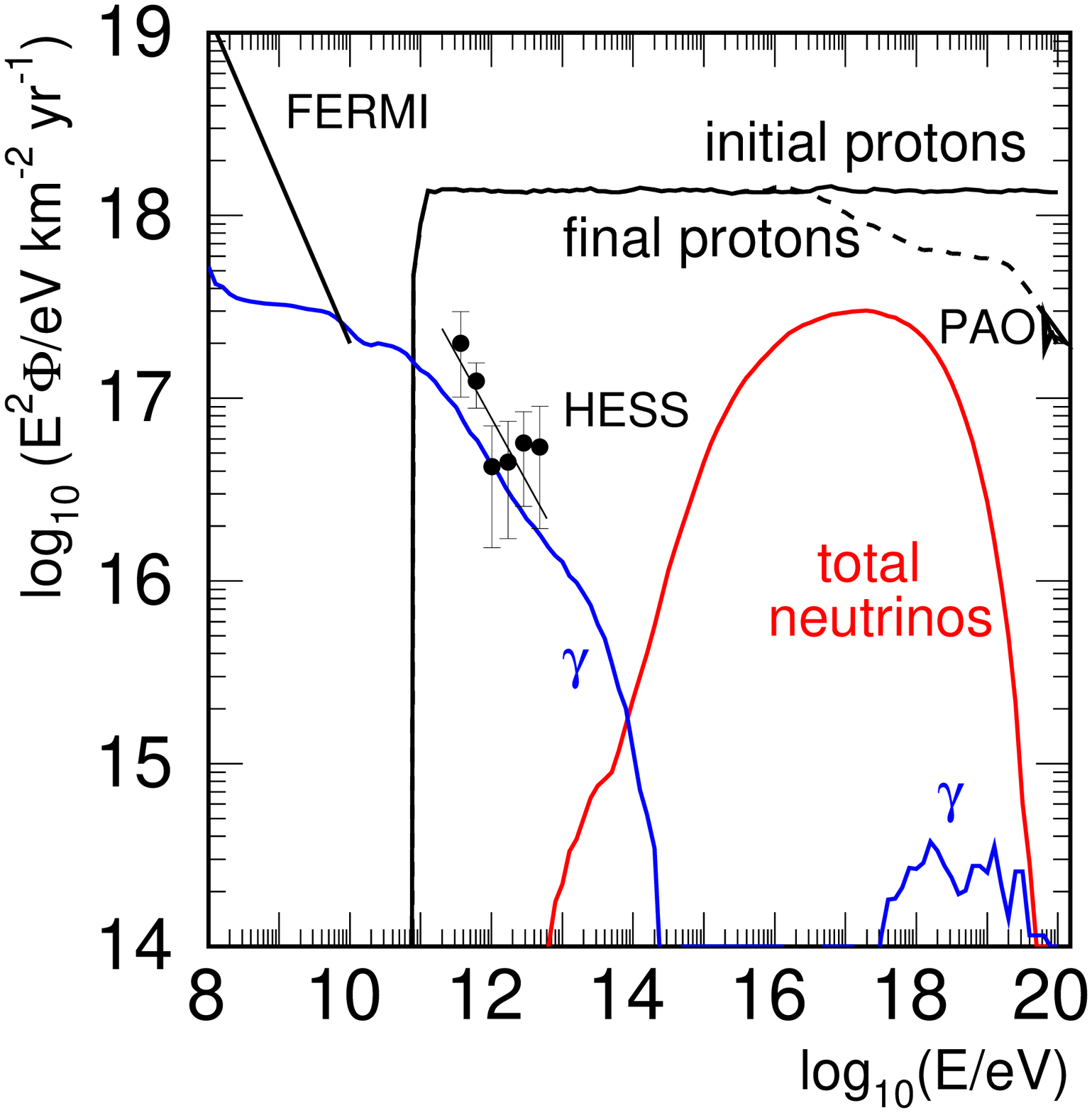,width=0.44\textwidth,angle=0}
\epsfig{file=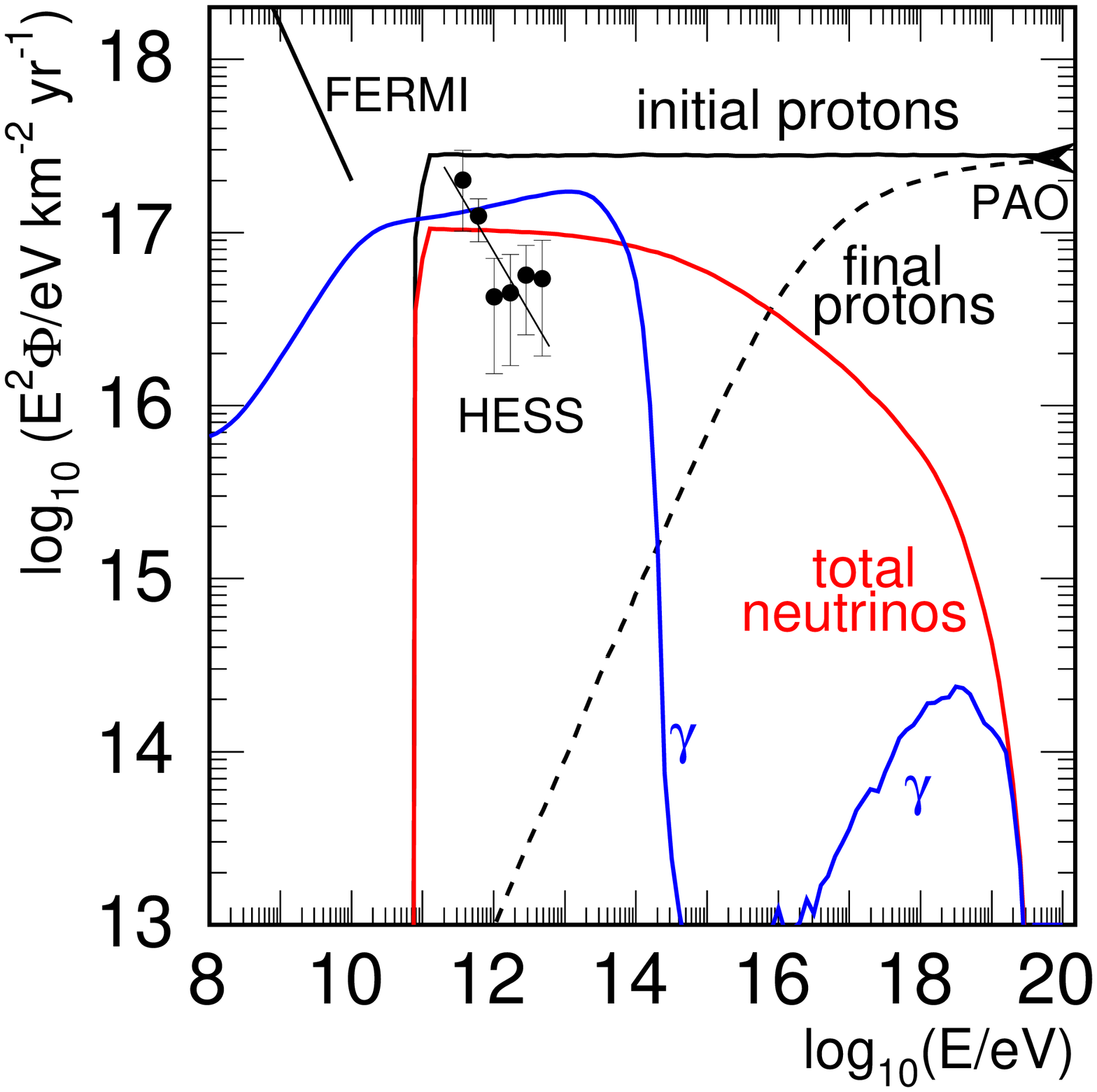,width=0.44\textwidth,angle=0}
\epsfig{file=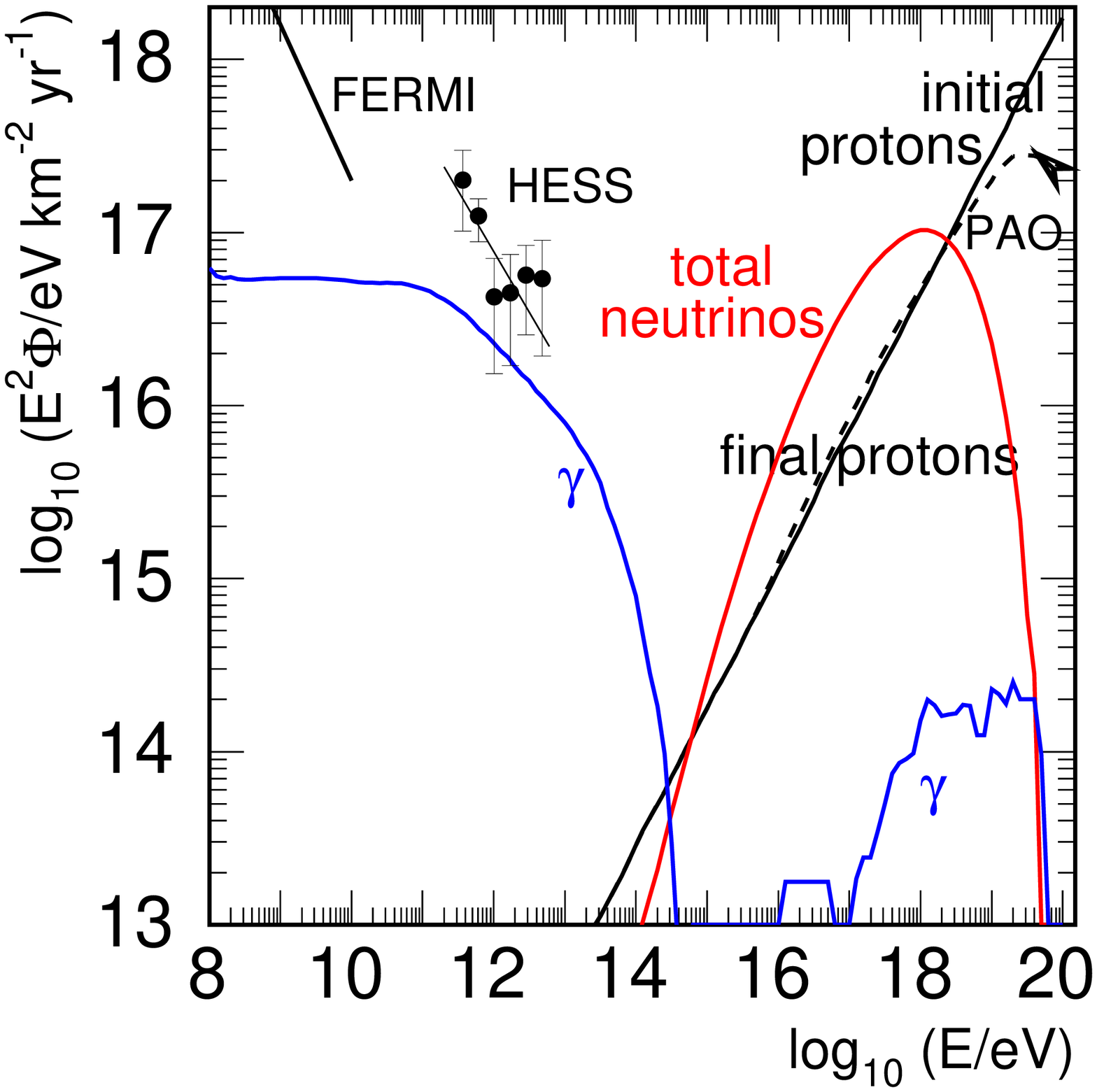,width=0.44\textwidth,angle=0}
\epsfig{file=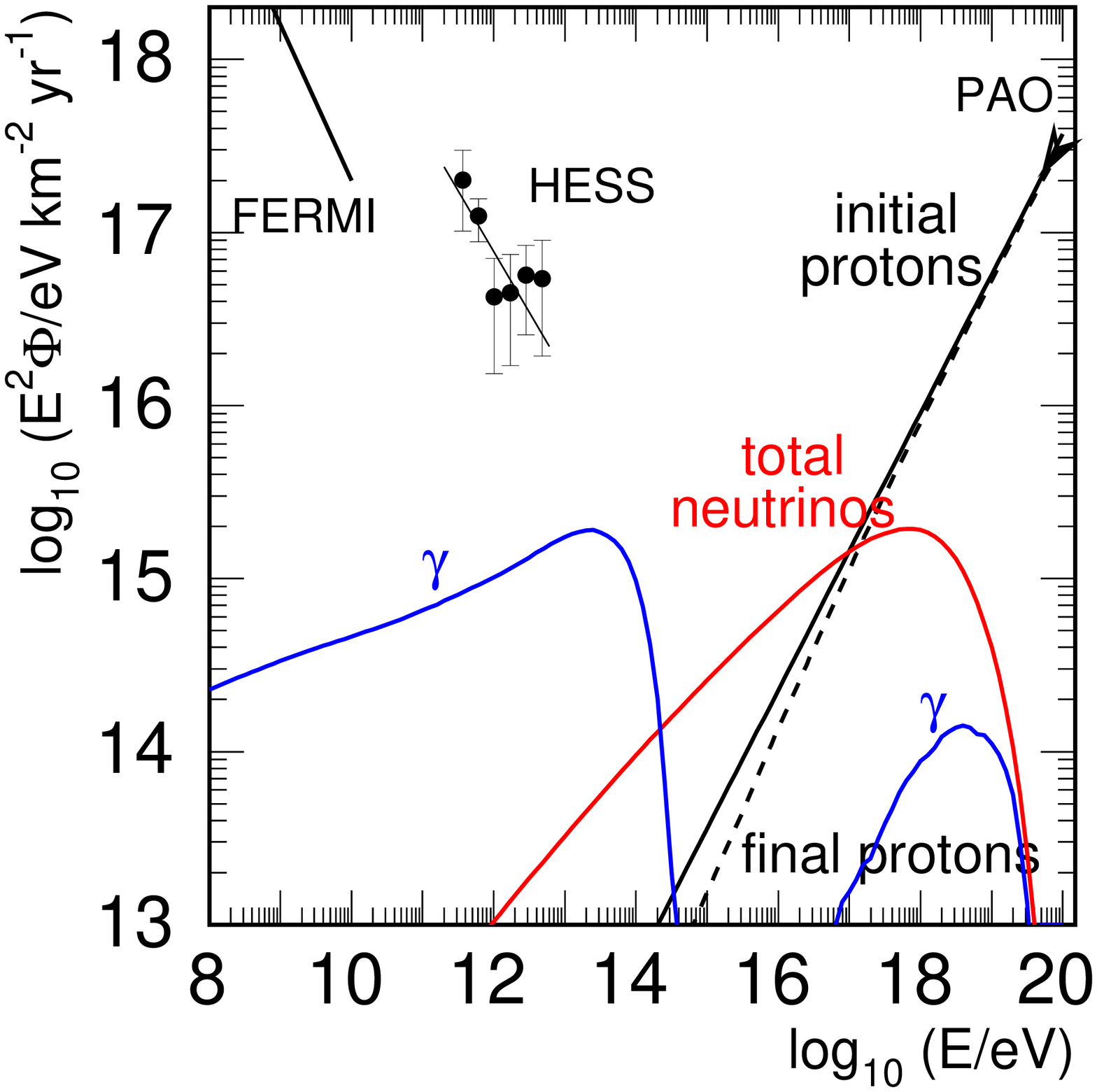,width=0.44\textwidth,angle=0}
\end{center}
\caption{Particle fluxes from Cen~A normalized to the PAO results, see the
text for a description.}
\label{spec1}
\end{figure}

Figure~\ref{spec1} displays the particle fluxes 
predicted\cite{Kachelriess:2008qx} from Cen~A
as function of the energy, assuming that the two events observed by
PAO around Cen~A indeed originate from this AGN. The case of
acceleration close to the core is shown on the left, while the case of
acceleration in the jet is shown on the right. From the top to the
bottom, spectra are displayed for a broken power-law, $\alpha=2$, and 
$\alpha=1.2$.
 In addition to the injected proton flux (black solid line),
we show the flux of protons (black dashed), photons (blue solid) and
neutrinos (red solid) arriving on Earth.  
Note that the cutoff in the neutrino and proton spectra below 100\,GeV
is artificial, since we neglect neutrinos and protons with lower
energies in our simulation.

In the core model the final proton flux is reduced by photon-proton
interactions by a factor $\approx 2$ above the threshold energy $\sim
10^{16}$\,eV (left), while diffusion in the jet increases the
interaction depth for lower energies (right), resulting in the
effective production of secondaries.
Since the CR spectra are normalized to the integral UHECR flux above
$E_{\rm th}= 5.6\times 10^{19}\,$eV, steeper spectra result in larger
secondary fluxes at low energies. 

While we could show earlier\cite{Kachelriess:2008qx} these fluxes only 
together with an upper limit from H.E.S.S.\ and the estimated sensitivity
of Fermi for point sources,  we have updated now these figures with the recently
published results from Fermi\cite{Abdo:2009wu} and
H.E.S.S.\\cite{Aharonian:2009xn}. Remarkably, the photon flux in the 
Fermi and the H.E.S.S.\ energy range has approximately the same 
power-law exponent ($\alpha\sim 2.6$ and 2.7), but the latter requires
a larger normalization constant. Such a behavior is expected, if a new
component, e.g.\ of hadronic origin, sets in above 100~GeV, while at lower
energies photons of electromagnetic origin dominate the spectrum.
For a more detailed test of this hypothesis in the future, the differential 
energy spectrum at the high-energy end of the Fermi spectrum will be most 
useful.

The most important consequence of the recent H.E.S.S.\ results is to
significantly disfavor the jet scenario. In spite of the uncertainties
in the normalization, the almost flat spectrum predicted in this model
is in contradiction with the shape of the $\gamma$-ray spectrum observed
by H.E.S.S.\ Moreover, the angular extension of the photon flux observed by 
H.E.S.S.\ is consistent with a point source at the AGN core and excludes
thereby the radio lobes as sources.

On the contrary, the shape of  the $\gamma$-ray spectrum observed
by H.E.S.S.\ agrees very 
well with the slope expected in the core model. The Fermi measurements
restrict additionally the source model, excluding the broken-power law 
case that leads to an excessive photon fluxes in the GeV range.
In summary, a primary
proton energy $dN/dE\propto E^{-\alpha}$ with $\alpha\sim 2$ or smaller
in the case of acceleration close to the core is consistent with the
H.E.S.S.\ and Fermi observations, assuming that in the GeV range
the photon flux has dominantly an electromagnetic origin.

Finally, we comment on the neutrino fluxes to be expected from Cen~A.
Calculating the expected event number in a neutrino telescope requires
a definite choice of the experiment.  Centaurus~A is from the location
of Icecube only visible from above, and thus the background of
atmospheric muons allows only the use of contained events that carry
essentially no directional information. 
By contrast, a neutrino telescope in the
Mediterranean could make use of the muon signal and the directional
information.
The resulting event numbers both for cascade and shower event number
per year observation time are summarized in Table~\ref{tableI}. 
For a discussion on the connection between Cen A as source of UHECRs
 and the associated diffuse neutrino flux see
Ref.~\refcite{Koers:2008hv}.

\begin{table}[ph]
\tbl{Number of neutrino events expected per year.}
{\begin{tabular}{|c||c|c|c|c|c|c|c|c|} 
\hline
 &\multicolumn{4}{|c|}{jet}&\multicolumn{4}{|c|}{core}\\
\cline{2-9}
$\alpha$ or $E_b$/eV & 1.2 & 2 & $10^{18}$ & $10^{17}$ 
         & 1.2 & 2&  $10^{18}$ & $10^{17}$ 
\\ \hline
contained $\# \;\nu$/yr & $8\times 10^{-5}$ & 0.02 & 0.4 & 2.0
                        & $7\times 10^{-4}$ & 0.01 & 0.3 & 0.9
\\  \hline
$\# \;\mu$/yr & $4\times 10^{-5}$ & 0.01 & 0.2 & 0.7
              & $3\times 10^{-4}$ & $7\times 10^{-3}$ & 0.1 & 0.5
\\ \hline
\end{tabular}
\label{tableI}}
\end{table}

\section{Conclusions}

In summary, the combination of data from PAO and H.E.S.S.\ is
consistent with a scenario where the UHECRs observed by PAO are
protons accelerated near the core. The VHE $\gamma$-rays detected by
H.E.S.S.\ would then have a hadronic origin, namely, the interaction of
the accelerated protons with the UV photons surrounding the core.
This model is expected to be tested in the near future as the
currently running PAO, H.E.S.S.\ and Fermi experiments increase their
statistics.


\section*{Acknowledgments}
S.O.\  acknowledges a Marie Curie IEF fellowship from the European Community,
R.T.\ support from the Deutsche Forschungsgemeinschaft within the SFB~676.


\end{document}